\documentclass[fleqn,twoside]{article}%
\usepackage{amsmath}
\usepackage{amsfonts}
\usepackage{amssymb}
\usepackage{graphicx}
\usepackage{caption}
\usepackage{cite}
\def\be{\begin{equation}}
\def\ee{\end{equation}}
\def\bes{\begin{equation}\begin{split}&}
\def\es{\end{split}}
\def\bi{\bibitem}

\begin{document}
\title{Perusing Buchbinder--Lyakhovich canonical formalism for Higher-Order Theories of Gravity.}
\author{Dalia Saha$^\dag$, Abhik Kumar Sanyal$^\ddag$}
\maketitle
\begin{center}
\noindent
$^{\dag}$ Dept. of Physics, University of Kalyani, West Bengal, India - 741235.\\
\noindent
$^{\dag},^{\ddag}$ Dept. of Physics, Jangipur College, Murshidabad,
\noindent
West Bengal, India - 742213. \\
\end{center}
\footnotetext[1] {\noindent
Electronic address:\\
\noindent
$^{\dag}$daliasahamandal1983@gmail.com\\
$^{\ddag}$sanyal\_ ak@yahoo.com}

\begin{abstract}
Ostrogradsky's, Dirac's and Horowitz's techniques of higher order theories of gravity produce identical phase-space structures. The problem is manifested in the case of Gauss-Bonnet-dilatonic coupled action in the presence of higher-order term, in which case, classical correspondence can't be established. Here, we explore yet another technique developed by Buchbinder and his collaborators (BL) long back and show that it also suffers from the same disease. However, expressing the action in terms of the three-space curvature, and removing ``the total derivative terms", if Horowitz's formalism or even Dirac's constraint analysis is pursued, all pathologies disappear. Here we show that the same is true for BL formalism, which appears to be the simplest of all the techniques, to handle.
\end{abstract}

\noindent
Keywords: Higher Order theory; Canonical Formulation.
\section{Introduction}

Canonical formulation of higher-order theories was developed by Ostrogradsky almost two centuries back \cite{1O, 2O}. However, it did not draw much attention, since other than toy mechanical models, practically no such physical theories were persuaded at that time. Exactly a century elapsed, when it was applied to a physically motivated problem, such as fourth order harmonic oscillator \cite{PU}. The real physical problem in this context appeared for the first time, while a renormalized quantum theory of gravity was attempted to formulate \cite{ST}. Higher-derivative theory of gravity is usually considered as a model of quantum gravity. The reason being, Einstein-Hilbert action is supplemented by curvature squared terms  ($R^2, ~R_{\mu\nu} R^{\mu\nu}$) to ensure renormalizability \cite{ST} and asymptotic freedom \cite{FT,T,BK}. Unfortunately, curvature-squared gravity theories have been found to suffer from the unresolved problem of physical unitarity in perturbative analysis, which is usual for higher-derivative theories. However, possibilities to overcome this difficulty were also discussed in some literatures \cite{T,AT} and references therein. It is also ascertained that curvature squared gravity would arise as a low-energy effective theory derived from superstring theory in $D = 10$ dimensions \cite{CH,N,Z}. Over the last couple of decades, higher order theories of gravity e.g., $F(R),~ F(\mathcal{G}),~F(R,T)$ etc, theories, ($R,~\mathcal{G}, ~T$ being the Ricci scalar, the Gauss-Bonnet term, and the torsion term respectively) have drawn much attention in search of alternatives to dark energy. Nonetheless, it is always suggestive to test viability of such modified theories of gravity in different contexts. In the context of the very early universe, a canonical formulation is required as a precursor, particularly to study quantum cosmology.\\

Since Ostrogradsky's technique does not apply in the degenerate case of singular Lagrangian, for which the Hessian determinant vanishes, Dirac's constraint analysis \cite{2D} may be applied for the purpose. Nonetheless, a host of theories have been formulated over decades to bypass the constraint analysis. One of these in this direction was originally proposed by Boulware \cite{Boul}, and later reformulated by Horowitz' \cite{3H}, in particular in the context of higher-order theory of gravity. Since the canonical formulation of higher order theories requires an extra degree of freedom, in Horowitz's formalism apart from the scale factor ($`a'$ in the Robertson-Walker minisuperspace) an auxiliary variable is introduced by taking derivative of the action (say $A$) with respect to the highest derivative of the field variable present ($Q = \frac{\partial A}{\partial \ddot a}$). In the end, the auxiliary variable is replaced by the basic variable (extrinsic curvature tensor) through a canonical transformation. The important finding in this regard is as follows: all the three formalisms, viz, Ostrogradsky's (once degeneracy has been removed), Dirac's and Horowitz's formalisms, produce an identical phase-space structure \cite{As11}. Meanwhile, certain pathologies with Horowitz' formalism have been identified. For example, it was noticed that Horowitz's formalism can even be applied in the case of linear gravity theory (Einstein-Hilbert action) leading to wrong quantum dynamics \cite{Po,As1,As2}, as well as some superfluous total derivative terms are eliminated \cite{As2,As3}, which neither may be obtained from the variational principle, nor having any connection with Gibbons-Hawking--York term \cite{Y, GH}, nor any of its modified versions, associated with higher-order gravity. Further, the coupling parameter, in the case of the ``non-minimally coupled scalar tensor theory of gravity associated with higher order term", has not been found to play any particular role, since its derivative does not appear in the Hamiltonian \cite{As13}. The same is true for the ``Dilatonic coupled Gauss-Bonnet-theory in the presence of higher order term", where additionally, the classical correspondence with quantum counterpart, could not be established \cite{As13}. In view of such an uncanny situation, yet another technique was developed, called the ``modified Horowitz's formalism" (MHF), which was successfully applied to different modified higher-order theories of gravity, to explore the evolution of the very early universe \cite{As1,As2,As3,As7,As8,As10,As11,As12,As13,As14,As15,As16,As18,As19,As20}. In the MHF, the action is expressed in terms of the three-space curvature (instead of the scale factor), ``the total derivative terms" are removed by integrating the action by parts, and Horowitz's formalism (the introduction of the auxiliary variable etc.) was followed, thereafter.\\

To be very specific, let us consider the following isotropic and homogeneous Robertson--Walker (RW) metric:

\be\label{RW} ds^2 = -N^2(t)~ dt^2 + a^2(t) \left[\frac{dr^2}{1-kr^2} + r^2 (d\theta^2 + sin^2 \theta d\phi^2)\right],\ee
for which the degeneracy in the Lagrangian disappears if the gauge ($N$) is fixed a priori, in which case, Ostrogradsky's technique applies as well. Once such degeneracy is removed, it is observed that Ostrogradsky's technique produces the same Hamiltonian, obtained following Horowitz's as well as Dirac's formalism \cite{As11}. Therefore, it certainly follows that both the Ostrogradsky's and Dirac's formalism implicitly suffer from the same problem, in disguise, as was noticed in Horowitz's technique, as discussed above. Therefore in the MHF, instead of the scale factor, the action is expressed in terms of the basic variable $h_{ij}$ --- the three space metric from the very beginning---so that redundant total derivative terms do not appear \cite{As2,As3}. Thereafter, all the total derivative terms are integrated out by parts, which become cancelled by the supplementary boundary (Gibbons--Hawking---York and modified Gibbons--Hawking---York) terms. Subsequently, the auxiliary variable is introduced following Horowitz's proposal. In the end, the auxiliary variable is replaced by the other basic variable $K_{ij}$ --- the extrinsic curvature tensor. In this process, the unwanted problems that appeared following Horowitz's formalism disappear, while it produces a different Hamiltonian altogether. We mention that although both Hamiltonians (obtained following the MHF and Ostrogradky's, Dirac's and Horowitz's formalisms) are related through the canonical transformation, they indeed produce different dynamics in the quantum domain. It is also important to mention that it is not possible to carry over the classical canonical transformations to the quantum domain for higher-order theories, due to the non-linearity. The MHF leads to an effective Hermitian Hamiltonian, a standard quantum mechanical probabilistic interpretation, and a viable semiclassical treatment, which exhibit oscillation of the wave function about the classical de-Sitter solution. As a result, the classical correspondence is established. In this regard, the MHF may be considered as the most-viable technique to handle the higher-order theories. It has later been established that, if the action is expressed in terms of the three-space metric ($h_{ij}$) from the very beginning and the total derivative terms are addressed, Dirac's constraint analysis \cite{2D} also produces the Hamiltonian identical to that of the MHF \cite{As13,As15,As16,As18}. \\

Amongst other techniques, Hawking-Luttrell technique \cite{HL} has limited application, since conformal transformation is not possible in general \cite{As3}, Schmidt's technique \cite{SC} is identical to the Horowitz's formalism in disguise \cite{As1}. However, there is yet another technique developed in the $80$'s by Buchbinder and his collaborators \cite{BL1,BL1.2, BL2, BL3, BL4}, which did not receive much attention. Querella \cite{Qphd} only noticed that although at a first glance, the general formalism developed by Buchbinder and his collaborators (BL) appears to be satisfactory, nevertheless it has pitfalls. BL formalism is our current concern. Here, we test this abstract theoretical settings of BL \textbf{formalism} in simple minisuperspace model to explore the pitfall, if any. The underlying essence of this formalism is to bypass Dirac's constrained analysis, very much like Horowitz's technique, but instead of introducing auxiliary variable, here the program is initiated with the basic variables $\{h_{ij},K_{ij}\}$, the three-space curvature and the extrinsic curvature tensors respectively from the very beginning. In our present attempt to explore the outcome of this technique, we discover that the formalism leads to identical phase space structure as was found in the case of Ostrogradsky's/Dirac's/Horowitz's formalism. \\

This paper is organized as follows. In the following section, we study scalar tensor theory of gravity (both the minimal and non-minimal cases), and Gauss-Bonnet-Dilatonic coupled action being supplemented by the scalar curvature squared ($R^2$) term, following BL formalism. In Section \ref{sec3}, we explore the fact that once total derivative terms are taken care of, the Hamiltonian does not differ from MHF. Section \ref{sec4} discusses its physical application, in connection with some earlier work. Section \ref{sec5}, concludes our work. \\

\section{BL Formalism in Three Different Higher Order Theories}

In view of the very importance of higher-order curvature invariant terms required to construct a renormalizable quantum theory of gravity when the curvature is extremely strong, a unique canonical formulation of the Einstein--Hilbert action being supplemented by higher-order curvature invariant terms, is therefore necessary. Here, we shall consider three different cases, minimally and non-minimally coupled scalar-tensor theory of gravity supplemented by $R^2$ term, and the scalar-tensor theory of gravity being supplemented by $R^2$ and Gauss-Bonnet terms. In the Robertson-Walker minisuperspace \eqref{RW} under consideration, the Ricci scalar and the Gauss-Bonnet terms are

\be \label{R1}R = \frac{6}{N^2}\left(\frac{\ddot a}{a}+\frac{\dot a^2}{a^2}+N^2\frac{k}{a^2}-\frac{\dot N\dot a}{N a}\right).\ee
\be \label{G1}\mathcal{G} = R^2 - 4R_{\mu\nu} R^{\mu\nu} + R_{\alpha\beta\mu\nu}R^{\alpha\beta\mu\nu}= \frac{24}{N^3 a^3}\left(N\ddot a - \dot N \dot a\right)\left(\frac{\dot a^2}{N^2} + k\right).\ee
respectively. For the sake of comparison with earlier results, we express actions in terms of the three space metric, instead of the scale factor, as its importance has been mentioned already, and will be explicitly shown at the beginning of Section \ref{sec3}. Since construction of higher-order theory to its canonical form requires an additional degree of freedom, hence, in addition to the three-space metric $h_{ij}$, the extrinsic curvature tensor $K_{ij}$ is treated as basic variable, as already stated. We therefore choose the basic variables $h_{ij} = z \delta_{ij}= a^2 \delta_{ij}$, so that $K_{ij} = -{\dot h_{ij}\over 2N} = -{a\dot a\over N} \delta_{ij}= -\frac{\dot z}{2 N} \delta_{ij}$. In terms of $z = a^2$, the Ricci scalar and the Gauss-Bonnet terms take the following forms,

\be \label{R2}R = {6\over N^2}\left[{\ddot z\over 2z} + N^2 {k\over z} - {1\over 2}{\dot N\dot z\over N z}\right],\ee

\be \label{G2}\mathcal{G}= {12\over N^2}\left({\ddot z\over z}-{\dot z^2\over 2z^2}-{\dot N\dot z\over  N z}\right)\left({\dot z^2\over 4 N^2z^2} + {k\over z}\right),\ee
respectively. It is noteworthy that since,

\be R_{\mu\nu}R^{\mu\nu} = {12\over N^4}\left[{\ddot a^2\over a^2}+{\dot a^2\ddot a\over a^3}+{\dot a^4\over a^4}-2{\dot N\dot a\ddot a\over N a^2}-{\dot N\dot a^3\over N a^3}+{\dot N^2 \dot a^2\over N^2 a^2}+k{N^2\ddot a\over a^3}+2k{N^2 \dot a^2\over a^4}-k{N \dot N\dot a\over  a^3}+k^2{N^4\over a^4}\right],\ee
therefore,

\be R_{\mu\nu}R^{\mu\nu} - {1\over 3} R^2 = -\left({12\over Na^3}\right){d\over dt}\left[{1\over 3}{\dot a^3\over N^3} + k {\dot a\over N}\right],\ee
and as a result,

\be \int\left[R_{\mu\nu}R^{\mu\nu} - {1\over 3} R^2\right]\sqrt{-g} d^4 x = -12C\int \left[{d\over dt}\left({1\over 3}{\dot a^3\over N^3} + k {\dot a\over N}\right)\right] dt\ee
is a total derivative term. Thus, $R_{\mu\nu}R^{\mu\nu}$ term is redundant in RW metric, once $R^2$ term is taken (the constant $C$ appears due to the integration of the three space). Hence, to scrutinize the BL formalism presented by Buchbinder and his collaborators in RW minisuperspace model \eqref{RW}, we consider scalar-tensor theories of gravity and also Gauss-Bonnet-Dilatonic coupled gravity theory, being associated with scalar curvature squared term $R^2$.

\subsection{Minimal coupling:}

Let us start with the following minimally coupled case,

\be\label{AA} A_1 = \int \sqrt{-g} \left[{\alpha}R + {\beta}R^2-\frac{1}{2}\phi_{,\mu}\phi^{,\nu}-V(\phi)\right]d^4x+\alpha\Sigma_{R} +\beta \Sigma_{R^2}.\ee
In the above, $\alpha = {1\over 16\pi G}$, $\beta$ is a constant coupling parameter, $\alpha \Sigma_{R}= 2\alpha\oint_{\partial\mathcal{V}}K \sqrt hd^3x$ is the Gibbons-Hawking--York boundary term \cite{GH} associated with Einstein--Hilbert sector of the above action, and $\beta\Sigma_{R^2} = 4\beta\oint_{\partial\mathcal{V}} R K\sqrt h d^3x$ is its modified version corresponding to $R^2$ term, while, $K$ is the trace of the extrinsic curvature tensor $K_{ij}$. Note that, both the counter terms are required under the condition $\delta R = 0$, at the boundary. Instead, if the condition $\delta K_{ij} = 0$ is chosen at the boundary, the counter terms are not required, as in the case of Horowitz's formalism \cite{3H}, since both the boundary terms appearing under metric variation vanish. However, in the case of Ostrogradsky's technique \cite{1O} and Dirac constrained analysis \cite{2D}, boundary terms are not taken care of. This is true for BL formalism too, as we shall see shortly. Nevertheless, the modified Horowitz's formalism \cite{As1, As2, As3, As7, As8, As10} fixes $\delta h_{ij}=0=\delta R$ at the boundary, and hence requires supplementary boundary terms. We have demonstrated earlier that proper attention to all the boundary terms is paid in modified Horowitz's formalism (MHF). As a result, it presents a different phase space structure of the Hamiltonian for a particular action being supplemented by higher-order terms. Nonetheless, it is related to the others under a suitable set of canonical transformation \cite{As11}. Although, as mentioned, such transformations cannot be carried over in the quantum domain, due to non-linearity. So, it's indeed required to check if BL formalism also produces the same. The action (\ref{AA}) in the RW minisuperspace model \eqref{RW} may be written in terms of the basic variable $h_{ij} = z \delta_{ij}$, as

\be\label{rr}\begin{split}
A_1 &= \int\Bigg[ {3\alpha\sqrt z}\Big(\frac{\ddot z}{ N}- \frac{\dot N \dot z}{N^2} + 2k N \Big) +\frac{9 \beta}{\sqrt z}\Big(\frac{{\ddot z}^2}{N^3} - \frac{2 \dot N \dot z \ddot z}{N^4} + \frac{{\dot N}^2{\dot z}^2}{N^5} -\frac{4k\dot N \dot z}{N^2}+ \frac{4 k {\ddot z}}{N} + 4 k^2 N \Big)\\& \hspace{1.0 cm}+z^{\frac{3}{2}}\Big(\frac{\dot\phi^2}{2N}-V N\Big)\Bigg]dt +\alpha \Sigma_R +\beta\Sigma_{R^2}.\end{split}\ee
The $(^0_0)$ component of the field equation in terms of the scale factor $`a'$ takes the following form

\be\label{A2}\begin{split}& \frac{6\alpha}{a^2}\left(\frac{\dot a^2}{N^2}+k\right)+\frac{36\beta}{a^2N^4}\Bigg(2\dot a\dddot a-2\dot a^2\frac{\ddot N}{N}-\ddot a^2-4\dot a\ddot a\frac{\dot N}{N}+2\dot a^2\frac{\ddot a}{a}+5\dot a^2\frac{\dot N^2}{N^2}-2\frac{\dot a^3\dot N}{a N}\\& \hspace{1.0 cm}-3\frac{\dot a^4}{a^2}-2kN^2\frac{\dot a^2}{a^2}+\frac{k^2N^4}{a^2}\Bigg)-\left(\frac{\dot\phi^2}{2N^2}+V\right)=0,\end{split}\ee
which contains term upto third derivative. This is the energy constraint equation ($E = 0$), and when expressed in terms of the phase space variables, becomes the Hamiltonian constraint equation, (due to diffeomorphic invariance) of the theory under consideration. This we aim at, following the formalism presented by Buchbinder and his collaborators (BL).\\

The action \eqref{AA} has already been expressed in terms of the basic variable $\{h_{ij}\}$, instead of the scale factor. Canonical formulation of higher order theories requires additional degree of freedom, and the only choice is the extrinsic curvature tensor $\{K_{ij}\}$. In contrast to Horowitz's formalism, where apart from $\{h_{ij}\}$ an auxiliary variable is introduced and at the end the Hamiltonian is expressed in terms of the basic variables $\{h_{ij},~K_{ij}\}$, in BL formalism, these basic variables are associated from the very beginning. In Robertson-Walker metric, the extrinsic curvature tensor is expressed as,

\be K_{ij} = -{\dot{h}_{ij}\over 2N} = -{2a\dot a\over 2N}\delta_{ij} = -\frac{\dot z}{2N}\delta_{ij} = -q_{ij} ~~\mathrm{say}. \ee
Since there is only one independent component, so instead of $q_{ij}$, the new generalized coordinate is chosen to be its trace viz,

\be\label{QBL} q = \frac{3\dot z}{2N}, \ \ \text{i.e.} \ \ q_{ij} = \frac{q}{3} \delta_{ij}.\ee
To express the action in terms of velocities, we choose,

\be v \equiv \dot q, \,\, v_{\phi}\equiv \dot\phi.\ee
The scalar curvature \eqref{R2} therefore takes the following form,

\be \label{R11}R = {2\dot q\over Nz} + \frac{6k}{z} \equiv R_q = {2\over N z}(v + 3N k),\ee
and action (\ref{rr}) can be expressed as,

\be\label{AQa} A_{1q}=\int\left[2\alpha\sqrt z(v + 3kN)+\frac{4\beta}{N\sqrt z}(v + 3kN)^2 +z^{\frac{3}{2}}\left(\frac{v_{\phi}^2}{2N}- N V\right)\right]dt,\ee
while the Lagrangian density is,

\be\label{LQ} L_{1q} = 2\alpha\sqrt z(v + 3kN)+\frac{4\beta}{N\sqrt z}(v + 3kN)^2 +z^{\frac{3}{2}}\left(\frac{v_{\phi}^2}{2N}- N V\right).\ee
Note that the boundary terms remain intact in the action as well as in the point Lagrangian. Canonical momenta are

\be\begin{split}\label{mom}p_q = \frac{\partial L_q}{\partial v} = 2\alpha \sqrt{z} + \frac{8\beta}{N\sqrt z}(v +3kN),~
p_N =\frac{\partial L_{1q}}{\partial v_N} = 0,~p_z=\frac{\partial L_q}{\partial v_z} = 0~ \text{and}~ p_{\phi}=\frac{\partial L_q}{\partial v_{\phi}}=\frac{z^{\frac{3}{2}}v_{\phi}}{N}.\end{split}\ee
Clearly, there exists two primary constraints $C \equiv p_N \approx 0$, and $D \equiv p_z \approx 0$. Therefore, Dirac constraint analysis appears to be essential. However, here is a wonderful twist \textbf{in} the BL formalism. For example, one can express the modified Lagrangian density as,

\be\label{LQ*}L_1^*=L_{1q}+p_q\left(\dot q-v\right)+p_N\left(\dot N-v_N\right)+p_z\left(\dot z-\frac{2Nq}{3}\right)+p_{\phi}\left(\dot\phi-v_{\phi}\right),\ee
and equivalently, the Hamiltonian density as,

\be H_1^* = p_q \dot q + p_N \dot N +p_\phi\dot\phi+ p_z \dot z - L_1^* = p_q v + p_N v_N+p_{\phi}v_{\phi} + p_z {2Nq\over 3} - L_{1q}.\ee
As a consequence, one can immediately find that the primary constraint $D \equiv p_z \approx 0$ disappears. Further, since $N$ is a non-dynamical Lagrange multiplier, hence the constraint $C$ vanishes strongly. Therefore, one arrives at,

\be H_1^* = p_q v + Cv_N +p_{\phi} v_\phi+ p_z{2qN\over 3} - L_{1q} = Cv_N+ p_q v + p_{\phi} v_\phi +  p_z {2qN\over 3} - L_{1q} = Cv_N + N {\mathcal H}^m_{BL}, \ee
where,

\be\begin{split} N\mathcal H^m_{BL} &= p_q v + p_{\phi} v_\phi  + {2N\over 3}q p_z - L_{1q} \\&= p_q v + p_{\phi} v_\phi + {2 N\over 3} q p_z - 2\alpha\sqrt{z}(v + 3kN) - {4\beta\over N\sqrt z}(v + 3kN)^2 - z^{3\over 2}\left({v_\phi^2\over 2N} + N V\right).\end{split}\ee
In the above, $m$ in the superscript stands for minimally coupled theory. Now upon substituting $v$ and $v_{\phi}$ from the definition of momentum \eqref{mom}, we obtain,

\be\label{bl} N\mathcal H^m_{BL} = N\left[\frac{2 q}{3}  p_z + \frac{\sqrt{z}}{16\beta}p_q^2 -\left(\frac{\alpha z}{4\beta}+3k\right)p_q +\frac{\alpha^2 }{4\beta} z^{\frac{3}{2}} +\frac{1}{2z^{\frac{3}{2}}}p_{\phi}^2 + Vz^{\frac{3}{2}}\right],\ee
so that the canonical Hamiltonian finally reads as,

\be\label{Hb1}\mathcal H^m_{BL}=\frac{2 q}{3}  p_z + \frac{\sqrt{z}}{16\beta}p_q^2 -\left(\frac{\alpha z}{4\beta}+3k\right)p_q +\frac{\alpha^2 }{4\beta} z^{\frac{3}{2}} +\frac{1}{2z^{\frac{3}{2}}}p_{\phi}^2 + Vz^{\frac{3}{2}}.\ee
The action (\ref{rr}) may also be cast in the canonical form,

\be\label{Aq1}\begin{split} A_{1q} =\int\left(\dot z p_z+\dot q p_q+\dot\phi p_{\phi} - N\mathcal H_{BL}\right)dt~ d^3 x\, \ \
= \int\left(\dot h_{ij} \pi^{ij} + \dot K_{ij}\Pi^{ij}+\dot\phi p_{\phi} - N\mathcal H_{BL}\right)dt~ d^3 x,\end{split}\ee
where, $\pi^{ij}$ and $\Pi^{ij}$ are momenta canonically conjugate to $h_{ij}$ and $K_{ij}$ respectively. For the sake of comparison, let us make the following canonical transformation:

\be\label{ct1} q \rightarrow {3\over 2}x;~~~~p_q\rightarrow {2\over 3}p_x,\ee
to express the above Hamiltonian (\ref{Hb1}) as:

\be\label{hlb} \mathcal H^m_{BL}=xp_z +\frac{\sqrt z}{36\beta}p_x^2-\left({\alpha z\over 6\beta} + 2k\right)p_x + \frac{\alpha^2z^{\frac{3}{2}}}{4\beta}+\frac{p_{\phi}^2}{2z^{\frac{3}{2}}}+Vz^{\frac{3}{2}}.\ee
It is revealed that the above Hamiltonian (\ref{hlb}) is exactly the one obtained earlier, following Ostrogradsky, Dirac as well as Horowitz's formalisms \cite{As11}. Note that, very much like the Ostrogradsky's and Dirac's formalisms here also, once the formalism is initiated, i.e. ($R$ is expressed in terms of $\{h_{ij}, K_{ij}\}$ \eqref{R11} as well as the action \eqref{AQa} and the point Lagrangian \eqref{LQ}), there remains no option to integrate the action by parts. As a result, even the Gibbons-Hawking--York term \cite{Y, GH} which is physically meaningful, being associated with the entropy of the black hole, along with its higher-order counterpart, also remain obscure. On the contrary, following Modified Horowitz's Formalism (MHF), where boundary terms are taken care of, we earlier obtained \cite{As11}

\be \label{M1} \mathcal H^m_{MHF} =  x p_z+\frac{\sqrt z}{36\beta}p_x^2+{3\alpha x^2\over 2\sqrt z}-{18\beta k x^2\over z^{3\over 2}}-\frac{36\beta k^2}{\sqrt z}-6k\alpha\sqrt z +\frac{p_{\phi}^2}{2z^{\frac{3}{2}}}+ Vz^{\frac{3}{2}}.\ee
Although the two (\ref{hlb}) and (\ref{M1}) exactly match under the following set of canonical transformations,

\[p_z \rightarrow p_z-\frac{18k\beta x}{z^{\frac{3}{2}}} + {3\alpha x\over 2\sqrt z},~~z \rightarrow z,\]
\[p_x \rightarrow p_x+\frac{36k\beta}{\sqrt z} -3\alpha\sqrt z,~~ x \rightarrow x,\]
\[p_{\phi} \rightarrow p_{\phi},~~ \phi \rightarrow \phi.\]
and apparently there is no contradiction between the two, note the essential difference: linear term in the momentum ($p_x$), which is very much present in (\ref{hlb}), remains absent from the Hamiltonian \eqref{M1}. As a result, the two Hamiltonians (\ref{hlb}) and (\ref{M1}) induce completely different quantum dynamics, since in the quantum domain, as mentioned, canonical transformation cannot be carried over due to non-linearity.

\subsection{Non-minimally coupled case}

We find that the two different Hamiltonians (\ref{hlb}) and \eqref{M1} render two different quantum descriptions of the same classical model. Although, some of the essential features (Gibbons-Hawking-York term and its higher order counterpart) are absent from the Hamiltonian (\ref{hlb}), it is not clear, which one gives correct quantum description of the theory. Further, there may exist a unitary transformation (we have not found it though) relating the two Hamiltonian operators. Therefore, to inspect the situation more deeply, we consider the non-minimally coupled case next, whose action

\be\label{Aa11} A_2 =\int\sqrt{-g}\;d^4x\left[f(\phi){R}+ \beta  R^2-\frac{1}{2}\phi_{,\mu}\phi^{,\nu}-V(\phi)\right]+f(\phi)\Sigma_{R} +B \Sigma_{R^2},\ee
may be expressed in the RW metric \eqref{RW} as

\be\label{A121}\begin{split}
A_2 &= \int\Bigg[ {3f(\phi)\sqrt z}\Big(\frac{\ddot z}{ N}- \frac{\dot N \dot z}{N^2} + 2k N \Big) +\frac{9 \beta}{\sqrt z}\Big(\frac{{\ddot z}^2}{N^3} - \frac{2 \dot N \dot z \ddot z}{N^4} + \frac{{\dot N}^2{\dot z}^2}{N^5} -\frac{4k\dot N \dot z}{N^2}+ \frac{4 k {\ddot z}}{N} + 4 k^2 N \Big)\\&\hspace{1.0 cm}+z^{\frac{3}{2}}\Big(\frac{\dot\phi^2}{2N}-V N\Big)\Bigg]dt +f(\phi) \Sigma_R +B\Sigma_{R^2},\end{split}\ee
where, as already mentioned, the supplementary boundary terms are required when MHF is taken into account. In the above, we consider an arbitrary functional coupling parameter $f(\phi)$. Pursuing the same procedure as above, one finally arrives at the following Hamiltonian:

\be\begin{split}\label{h2b}& {\mathcal H^{nm}}_{BL}=xp_z+\frac{\sqrt{z}}{36\beta}p_x^2-\left(f(\phi)\frac{z}{6\beta} + 2k\right)p_x + f^2(\phi)\frac{z^{\frac{3}{2}}}{4\beta}+\frac{p_{\phi}^2}{2z^{\frac{3}{2}}}+Vz^{\frac{3}{2}},\end{split}\ee
which is again identical to the one found following Dirac formalism and may be found following Ostrogradsky's and Horowitz's techniques as well \cite{As13}. In the superscript $nm$ stands for non-minimal coupling. The action (\ref{A121}) may also be cast in the canonical form as in \eqref{Aq1}. On the contrary, following MHF, one finds \cite{As13}

\be\begin{split}\label{M2} {\mathcal H^{nm}}_{MHF}&= x p_z+{\sqrt z\over 36\beta}{p_x^2}+3f(\phi)\left({x^2\over 2\sqrt z}-2k\sqrt z\right)-{18k\beta\over \sqrt z}\left({x^2\over z}+2k \right)+{p_{\phi}^2\over 2z^{3\over 2}}\\&+{3x f'(\phi)p_{\phi}\over z}+{9f'(\phi)^2x^2\over 2\sqrt z}+ V z^{3\over 2}.\end{split}\ee
However, under the following set of canonical transformations,

\[p_z \rightarrow p_z - \frac{18\beta k x}{z^{\frac{3}{2}}} + {3f(\phi) x\over 2\sqrt z},~~ z\rightarrow z,\]
\[p_x \rightarrow p_x+36 \beta {k\over \sqrt z} + 3f(\phi)\sqrt z, ~~x \rightarrow x,\]
\[p_{\phi} \rightarrow p_{\phi}+3 f'(\phi){x\sqrt z}, ~~ \phi\rightarrow \phi,\]
the two Hamiltonians \eqref{h2b} and \eqref{M2} match again \cite{As13}. Nevertheless, here the difference is predominant and explicit. Note that $f'(\phi)$ term does not appear in \eqref{h2b}, while it is coupled to $p_\phi$ in \eqref{M2}. This coupled ($f'(\phi) p_\phi$) term requires operator ordering in the quantum domain, which is different for different form of $f(\phi)$. Hence even if the two Hamiltonians are related through unitary transformation, such transformation would be different for different form of $f(\phi)$.

\subsection{Einstein-Gauss-Bonnet-Dilatonic action in the presence of higher order term}

Although, it is clear that two different quantum descriptions follow from the same classical action using different techniques, it is still abstruse to select the correct description. Therefore, we next consider Einstein-Gauss-Bonnet-Dilatonic coupled action in the presence of higher order curvature invariant term. Gauss-Bonnet (GB) term arises quite naturally as the leading order of the $\alpha'$ expansion of heterotic superstring theory, where $\alpha'$ is the inverse string tension \cite{GB0,GB1,GB2,GB3,GB3a,GB4}. Several interesting features of GB term have been explored in the past and appear in the literature \cite{GB5,GB6,GB7,GB8,GB9,GB10,GB11,GB12,GB13,GB14,GB15,GB16,GB17,GB18,GB19,GB20,GB21,GB22,GB23,GB24,GB25}. However, Gauss–Bonnet term is topological invariant in $4$-dimensions, and so to get its contribution in the field equations, a dynamic dilatonic scalar coupling is required. It is worth mentioning  that, in string induced gravity near initial singularity, GB coupling with scalar field plays a very crucial role for the occurrence of nonsingular cosmology \cite{GB26,GB27}. The particular hallmark of GB term is the fact that, despite being formed from a combination of higher order curvature invariant terms $(\mathcal {G} = R^2 - 4R_{\mu\nu}R^{\mu\nu} + R_{\alpha\beta\mu\nu}R^{\alpha\beta\mu\nu})$ \eqref{G1}, it ends up only with second order field equations, avoiding Ostrogradsky's instability, and equivalently, ghost degrees of freedom. Nonetheless, such a wonderful feature ultimately leads to a serious pathology of `Branched Hamiltonian', which has no unique resolution till date \cite{BH1,BH2,BH3}. Nevertheless, it has been revealed that, the pathology may be bypassed upon supplementing the action with higher order curvature invariant term \cite{As8,As10}. We therefore consider the following action,

\be\label{Aa1} A_3 =\int\sqrt{-g}\;d^4x\left[\alpha{R}+\beta R^2+\gamma(\phi)\mathcal{G}-\frac{1}{2}\phi_{,\mu}\phi^{,\nu}-V(\phi)\right] + \alpha\Sigma_R + \beta\Sigma_{R^2} + \gamma(\phi)\Sigma_\mathcal{G}.\ee
In the above, Gauss-Bonnet term $\mathcal{G}$, is coupled with $\gamma(\phi)$, while $V(\phi)$ is the dilatonic potential. Further, the symbol $\mathcal{K}$ stands for $\mathcal{K}=K^3 - 3K K^{ij}K_{ij} + 2K^{ij}K_{ik}K^k_j$, where, $K$ is the trace of the extrinsic curvature tensor $K_{ij}$, and $\gamma(\phi)\Sigma_\mathcal{G} = 4\gamma(\phi)\oint_{\partial\mathcal{V}} \left( 2G_{ij}K^{ij} + \frac{\mathcal{K}}{3}\right)\sqrt hd^3x$ is the supplementary boundary term associated with Gauss-Bonnet sector. The $(^0_0)$ component of the Einstein's field equation in terms of the scale factor here reads as,

\be\begin{split}\label{A22}& \frac{6\alpha}{a^2}\left(\frac{\dot a^2}{N^2}+k\right)+\frac{36\beta}{a^2N^4}\Bigg(2\dot a\dddot a-2\dot a^2\frac{\ddot N}{N}-\ddot a^2-4\dot a\ddot a\frac{\dot N}{N}+2\dot a^2\frac{\ddot a}{a}+5\dot a^2\frac{\dot N^2}{N^2}-2\frac{\dot a^3\dot N}{a N}\\& \hspace{1.0 cm}-3\frac{\dot a^4}{a^2}-2kN^2\frac{\dot a^2}{a^2}+\frac{k^2N^4}{a^2}\Bigg)+\frac{24\gamma'\dot a\dot\phi}{N^2a^3}\left(\frac{\dot a^2}{N^2}+k\right) - \left(\frac{\dot\phi^2}{2N^2}+V\right)=0.\end{split}\ee
The action (\ref{Aa1}) in terms of the basic variable ($h_{ij} = a^2\delta_{ij} = z\delta_{ij}$) may be expressed as,

\be\begin{split}\label{A12}
A_3 &= \int\Big[{3\alpha\sqrt z}\left(\frac{\ddot z}{ N}- \frac{\dot N \dot z}{N^2} + 2k N \right)+\frac{9\beta}{\sqrt z}\left(\frac{{\ddot z}^2}{N^3} - \frac{2 \dot N \dot z \ddot z}{N^4} + \frac{{\dot N}^2{\dot z}^2}{N^5} -\frac{4k\dot N \dot z}{N^2}+ \frac{4 k {\ddot z}}{N} + 4 k^2 N\right) \\& \hspace{1.0 cm}+ \frac {3 \gamma (\phi)}{N \sqrt z}\left(\frac{{\dot z}^2 \ddot z}{N^2 z} + 4 k \ddot z - \frac{{\dot z}^4}{ 2 N^2 z^2} - \frac{\dot N {\dot z}^3}{N^3 z} - \frac{2 k {\dot z}^2}{z}- \frac{4 k \dot N \dot z}{N}\right)+z^{\frac{3}{2}}\left(\frac{1}{2N}\dot\phi^2-VN\right)\Big]dt\\&\hspace{1.0 cm}+\alpha\Sigma_R+ \beta\Sigma_{R^2} + \gamma(\phi)\Sigma_\mathcal{G},
\end{split}\ee
where the additional supplementary boundary term $\gamma(\phi)\Sigma_\mathcal{G} = -\gamma(\phi) \frac {\dot z}{N \sqrt z}\Big(\frac{{\dot z}^2}{N^2 z} + 12 k\Big)$, is required in the case of MHF. Inserting the other basic variable ($K_{ij} = -{q\over 3}\delta _{ij}$) and considering $\dot q = v$ \eqref{QBL}, the action (\ref{A12}) finally may be expressed as,

\be\begin{split}\label{LQq} A_{3q} &=\int\Bigg[2\alpha\sqrt z(v + 3kN)+\frac{4\beta}{N\sqrt z}(v + 3kN)^2 +\frac{8\gamma(\phi)}{\sqrt z}\left[{{v}q^2\over 9z}-{q^4 N\over 27 z^2}+{kv}-{kNq^2\over 3z}\right]\\& \hspace{1.0 cm} +z^{\frac{3}{2}}\left(\frac{v_{\phi}^2}{2N}- N V\right)\Bigg]dt+ \alpha\Sigma_R+ \beta\Sigma_{R^2} + \Lambda(\phi)\Sigma_\mathcal{G}.\end{split}\ee
Thus, the Lagrangian density takes the following form,

\be\begin{split}\label{LQ1} L_{3q}&=2\alpha\sqrt z(v + 3kN)+\frac{4\beta}{N\sqrt z}(v + 3kN)^2+\frac{8\gamma(\phi)}{\sqrt z}\left[{vq^2\over 9z}-{q^4 N\over 27 z^2}+{kv}-{kNq^2\over 3z}\right]+z^{\frac{3}{2}}\left(\frac{1}{2N}v_\phi^2-VN\right),\end{split}\ee
where, boundary terms are not taken care of. The canonical momenta are

\be \label{Mom0}\begin{split} & p_q = \frac{\partial L_q}{\partial v} = 2\alpha\sqrt z + \frac{8\beta}{N\sqrt z}\left({v}+3kN\right)+{8\gamma(\phi)\over{\sqrt z}}\left(\frac{q^2}{9z}+k\right),\\&
p_N = \frac{\partial L_{3q}}{\partial v_N} =0,~~p_\phi = \frac{\partial L_q}{\partial v_\phi}=\frac{z^{\frac{3}{2}}v_{\phi}}{N},~~\text{and}, ~~ p_z = \frac{\partial L_{3q}}{\partial v_z} = 0.\end{split}\ee
Clearly, there exists two primary constraints $C \equiv p_N \approx 0$, and $D \equiv p_z \approx 0$, which are usually handled by Dirac constraint analysis. However, as mentioned, such analysis is not at all required in the BL formalism. For example, one can express the modified Lagrangian density as,

\be\label{LQ1*}L_3^*=L_{3q}+p_q\left(\dot q-v\right)+p_N\left(\dot N-v_N\right)+p_z\left(\dot z-\frac{2Nq}{3}\right)+p_{\phi}\left(\dot\phi-v_{\phi}\right),\ee
so that the corresponding Hamiltonian density takes the following form,

\be H_3^* = p_q \dot q + p_N \dot N +p_\phi\dot\phi+ p_z \dot z - L_3^* = p_q v + p_N v_N+p_{\phi}v_{\phi} + {2Nq\over 3}p_z  - L_{3q}.\ee
As a result, the primary constraint $D \equiv p_z \approx 0$ disappears and one obtains,

\be H_3^* = p_q v + Cv_N +p_{\phi} v_\phi+ p_z{2qN\over 3} - L_{3q} = Cv_N+\left(p_q v +p_{\phi} v_\phi + {2qN\over 3}p_z  - L_{3q}\right) = Cv_N+ N {\mathcal H^{GB}}_{BL}. \ee
In the superscript $GB$ stands for Hamiltonian in connection with Einstein-Gauss-Bonnet-Dilatonic coupling. Note that the constraint $C \equiv p_N$ strongly vanishes, since the lapse function $N$ is simply a Lagrange multiplier. Therefore,

\be\begin{split} N{\mathcal H^{GB}}_{BL} &= p_q v + p_{\phi}v_{\phi}+ {2qN\over 3}p_z - L_{3q}
\\&= p_q v +p_{\phi}v_{\phi}+ {2qN\over 3}p_z - 2\alpha\sqrt z(v + 3kN)-\frac{4\beta}{N\sqrt z}(v + 3kN)^2
\\&\hspace{1.0 cm}-\frac{8\gamma(\phi)}{\sqrt z}\left[{vq^2\over 9z}-{q^4 N\over 27 z^2}+{k v}-{kNq^2\over 3z}\right] -z^{\frac{3}{2}}\left(\frac{1}{2N}v_\phi^2-VN\right).\end{split}\ee
Now upon substituting $v$ from the definition of momentum \eqref{Mom0}, one obtains,

\be\begin{split}\label{BLc} N{\mathcal H^{GB}}_{BL}=& N\Big[\frac{2qp_z}{3}+\frac{{\sqrt z}p_q^2}{16\beta}-p_q\left(\frac{\alpha z}{4\beta}+3k\right)+\frac{\alpha^2 z^{3\over 2}}{4\beta}-p_q\left(\frac{\gamma q^2}{9\beta z}+\frac{\gamma k}{\beta}\right)+\frac{2\alpha\gamma}{\beta}\left(\frac{q^2}{9\sqrt z}+k{\sqrt z}\right)\\&\hspace{1.0 cm}+{4\gamma q^4\over 27 z^{3\over 2}}\left(\frac{\gamma}{3\beta}+2\right)+\frac{8\gamma kq^2}{3z^{3\over 2}}\left(\frac{\gamma}{3\beta}+2\right)+\frac{12\gamma k^2}{\sqrt z}\left(\frac{\gamma}{3\beta}+2\right)+\frac{p_{\phi}^2}{2z^{\frac{3}{2}}}+Vz^{\frac{3}{2}}\Big].\end{split}\ee
The canonical Hamiltonian therefore finally reads as,

\be\begin{split}\label{HBL1}{\mathcal H^{GB}}_{BL}=&\frac{2qp_z}{3}+\frac{{\sqrt z}p_q^2}{16\beta}-p_q\left(\frac{\alpha z}{4\beta}+3k\right)+\frac{\alpha^2 z^{3\over 2}}{4\beta}-p_q\left(\frac{\gamma q^2}{9\beta z}+\frac{\gamma k}{\beta}\right)+\frac{2\alpha\gamma}{\beta}\left(\frac{q^2}{9\sqrt z}+k{\sqrt z}\right)\\&\hspace{1.0 cm}+{4\gamma q^4\over 27 z^{3\over 2}}\left(\frac{\gamma}{3\beta}+2\right)+\frac{8\gamma kq^2}{3z^{3\over 2}}\left(\frac{\gamma}{3\beta}+2\right)+\frac{12\gamma k^2}{\sqrt z}\left(\frac{\gamma}{3\beta}+2\right)+\frac{p_{\phi}^2}{2z^{\frac{3}{2}}}+Vz^{\frac{3}{2}}.\end{split}\ee
Again, for the sake of comparison, let us make the canonical transformation $q \rightarrow {3\over 2}x;~p_q\rightarrow {2\over 3}p_x$ \eqref{ct1}, to express the above Hamiltonian (\ref{HBL1}) in the following form,

\be\begin{split}\label{H1BL} {\mathcal H^{GB}}_{BL}&=x p_z+\frac{\sqrt{z}p_x^2}{36\beta}+\frac{\alpha^2 z^{\frac{3}{2}}}{4\beta}-\left(\frac{\alpha z }{6\beta}+\frac{\gamma x^2}{6\beta z}+\frac{2k\gamma}{3\beta}+2k\right)p_x +\frac{p_{\phi}^2}{2z^{\frac{3}{2}}}+\left(\frac{\gamma^2}{4\beta z^{5\over 2}}+\frac{3\gamma}{2 z^{5\over 2}}\right)x^4\\&\hspace{1.0 cm}+\left(\frac{\alpha\gamma}{2\beta \sqrt z}+\frac{12k\gamma}{z^{3\over 2}}+\frac{2k\gamma^2}{\beta z^{3\over 2}}\right)x^2+\frac{2\alpha k\gamma\sqrt z}{\beta}+\frac{24k^2\gamma}{\sqrt z}+\frac{4k^2\gamma^2}{\beta\sqrt z} +Vz^{\frac{3}{2}},\end{split}\ee
and notice that, it is similar to the one already found, following Dirac formalism and may be found following Ostrogradsky's and Horowitz's techniques as well \cite{As13}. The action (\ref{A12}) may also be cast in the canonical form with respect to the basic variables as,

\be\label{Aq}\begin{split} A_{3q} &=\int\left(\dot z p_z+\dot q p_q+\dot\phi v_{\phi} - N\mathcal {H}_{BL}\right)dt~ d^3 x
= \int\left(\dot h_{ij} \pi^{ij} + \dot K_{ij}\Pi^{ij}+\dot\phi v_{\phi} - N\mathcal{H}_{MHF}\right)dt~ d^3 x,\end{split}\ee
where, $\pi^{ij}$ and $\Pi^{ij}$ are momenta canonically conjugate to $h_{ij}$ and $K_{ij}$ respectively. Hence, everything appears to be consistent. On the contrary, although following MHF, we found \cite{As13}

\be\begin{split}\label{M3} {\mathcal H^{GB}}_{MHF}&= x p_z+\frac{\sqrt{z}p_x^2}{36\beta}+{3\alpha}\left(\frac{x^2}{2\sqrt z}-2k\sqrt z \right)-{18k\beta\over \sqrt z}\left({x^2\over z}+2k \right)+\left({x^6\over 2z^{9\over 2}}+{12kx^4\over z^{7\over 2}}+{72k^2x^2\over z^{5\over 2}} \right)\gamma'^2 \\&\hspace{1.0 cm}+\left({x^3\over z^3}+{12kx\over z^2} \right)\gamma'p_{\phi}  +\frac{p_{\phi}^2}{2Z^{\frac{3}{2}}}+Vz^{\frac{3}{2}},\end{split}\ee
nonetheless, under the following set of canonical transformations,

\be\begin{split}&p_z \rightarrow p_z - \frac{18\beta k x}{z^{\frac{3}{2}}} + {3\alpha x\over 2\sqrt z}-\frac{6k\gamma(\phi) x} {z^{\frac{3}{2}}}-\frac{3\gamma(\phi) x^3}{2z^{\frac{5}{2}}},~~ z\rightarrow z,\\&
p_x \rightarrow p_x+36 \beta {k\over \sqrt z} + 3\alpha\sqrt z+ {3\gamma(\phi) x^2\over z^{3\over 2}}+{12k\Lambda \over \sqrt z}, ~~x \rightarrow x,\\&
p_{\phi}\rightarrow p_{\phi}-{\gamma'(\phi) x^3\over z^{3\over 2} }-{12k\gamma'(\phi) x\over \sqrt z},  ~~ \phi\rightarrow \phi,\end{split}\ee
the two Hamiltonians \eqref{H1BL} and \eqref{M3} match again \cite{As13}. Apparently therefore, there is absolutely no problem. Nevertheless note that, the Hamiltonian \eqref{M3} contains a term ($\gamma'(\phi)p_\phi$), which is absent from \eqref{H1BL}. Now, during canonical quantization the presence of this term requires operator ordering, which is different for different form of $\gamma(\phi)$. As a result, even if the two may be related through unitary transformation, such transformation would be different for different form of $\gamma(\phi)$. Thus, there does not exist a unique unitary transformation. In a nutshell, we repeat that the two Hamiltonians \eqref{H1BL} and \eqref{M3} induce two different descriptions in the quantum domain, and apparently, there is no way to choose one to be the correct.

\section{The role of divergent terms:}

The very first important point to mention is, in all the formalisms the scale factor is treated as the basic variable, while we initiate our program treating three three-space curvature, instead. To explain the reason behind this choice, let us consider curvature squared action, $A = \int\beta R^2 d^4x$, as an example. Under variation, it gives a total derivative term $\sigma = -4\beta \int R K \sqrt{h}~d^3x$, as mentioned earlier, where $K$ is the trace of the extrinsic curvature tensor $K_{ij}$. A counter term ($-\sigma$), known by the name modified Gibbons-Hawking--York term \cite{Y,GH}, must be added to the action in case, instead of $\delta \dot q$, $\delta R$ is kept fixed at the boundary, as in MHF. In the RW \eqref{RW} metric under consideration, the action reads as,

\be A = 36\beta \int \left[a\ddot a^2 +2\dot a^2\ddot a+2k\ddot a + {\dot a^4\over a} + {2k\dot a^2\over a} + {k^2\over a}\right]dt\int d^3 x.\ee
Under integration by parts, we end up with,

\be A = C\int\left[a\ddot a^2 + {\dot a^4\over a} + {2k\dot a^2\over a} + {k^2\over a}\right]dt + C\left({2\over 3}\dot a^3+2k\dot a\right).\ee
where, $C = 36\beta\int d^3 x$. Now following Horowitz's program, we introduce an auxiliary variable $Q = {\partial A\over \partial \ddot a} = 2Ca \ddot a$, judiciously into the action in the following manner, such that it may be cast in canonical form,

\be A = \int \left[Q\ddot a - {Q^2\over 4 C a} +C\left({\dot a^4\over a} + {2k\dot a^2\over a} + {k^2\over a}\right)\right]dt+ C\left({2\over 3}\dot a^3+2k\dot a\right).\ee
Integrating the action again by parts we find

\be A = \left[-\dot Q\dot a- {Q^2\over 4 C a} +C\left({\dot a^4\over a} + {2k\dot a^2\over a} + {k^2\over a}\right)\right]+ C\left({Q\dot a\over C}+{2\over 3}\dot a^3+2k\dot a\right).\ee
The action is canonical, since the Hessian determinant is non-zero. It is trivial to check that the above action gives correct field equations, but the left out total derivative term may be expressed as,

\be \sigma' = -4\beta \int R K \sqrt{h}~d^3x + 16\beta \int K\sqrt{h} \left({\dot a^2\over a^2}\right) d^3x,\ee
and as a result $\sigma \ne \sigma'$. Thus some redundant total derivative terms are pulled out in the process, which has severe consequence in the quantum domain. On the contrary, if we start with, $z = a^2$, the action reads as,

\be A = C\int\left({\ddot z^2\over 4\sqrt z} + {k\ddot z\over \sqrt z} + {k^2\over \sqrt z}\right)dt = C\int \left({\ddot z^2\over 4\sqrt z} + {^2\over \sqrt z} \right) + C{k\dot z\over \sqrt z},\ee
where the last expression is found under integration by parts. Now following Horowitz's program, we find the auxiliary variable as $Q = {\partial A\over\partial \ddot z} = C{\ddot z\over 2\sqrt z}$, which is again judiciously introduced in the action as,

\be A = \int \left[Q\ddot z - {\sqrt z Q^2\over C} + C\left({k\ddot z\over \sqrt z} + {k^2\over \sqrt z}\right)\right] dt +  C{k\dot z\over \sqrt z}.\ee
Finally, performing integration by parts again, one obtains,

\be A = \int \left[-\dot Q\dot z - {\sqrt z Q^2\over C} + C\left({k\ddot z\over \sqrt z} + {k^2\over \sqrt z}\right)\right] dt +  C\left[{Q\dot z\over C} + {k\dot z\over \sqrt z}\right],\ee
The action is again canonical, the Euler-Lagrange equations here again lead to the appropriate field equations, while one can express the total derivative term as $\sigma$. In a nut-shell, although total derivative terms do not affect the classical field equations, for non-linear theories such as gravity, such terms tell upon the quantum dynamics. Therefore, to establish consistency in every respect, $h_{ij}$ should be treated as the basic variable, instead of the scale factor. This is essentially the so-called MHF, which finally requires to replace the auxiliary variable by the the second basic variable, viz., the extrinsic curvature tensor $K_{ij} = -a\dot a = - \dot z = x (\mathrm{say})$, in the Hamiltonian.\\

Next, we observe that the phase-space structures obtained following BL formalism although are identical to the Ostrogradsky/Dirac/Horowitz's formalism, they all differ from the MHF upto a canonical transformation. We quote from \cite{As13} the general argument in connection with the total derivative terms, which runs as; ``it is just the change of the variables in the wave function and the phase transformation, plus the change of the integration measure, and the transformation of the momenta respecting the change of the measure, and so a unitary transformation relates the two". It's possible (we have not found though) that each pair of quantum equations cast from \{(\ref{hlb}) and \eqref{M1}\}; \{\eqref{h2b} and \eqref{M2}\}; \{\eqref{H1BL} and \eqref{M3}\}, are related by unitary transformation. However, it was also mentioned \cite{As13} that different forms of coupling parameter yield different quantum dynamics in the case of MHF, due to the presence of a coupling term ($f'(\phi)p_\phi$) for non-minimal coupled case, and ($\gamma'(\phi) p_\phi$) for the Gauss-Bonnet-Dilaton coupled case, in the Hamiltonian. Thus, different unitary transformations (if exist) are required to relate the last two pairs. Such coupling as well as the derivative of coupling parameter remain absent in other formalisms. In a nutshell, unitary transformation relating each pair is not unique. Further, the semiclassical wave functions found for all the three cases studied here, exhibit different pre-factors and exponents for each pair \cite{As13}. This generates different probability amplitude and the evolution of the wave function while entering the classical domain. \\

Finally, it is important to note that, if the coupling parameter $f(\phi)$ is treated as constant in Subsection \ref{sec2.2}, the Hamiltonian \eqref{M2} merely reduces to \eqref{M1}, while the Hamiltonian \eqref{h2b} reduces to (\ref{hlb}). Hence the question is: which of the two should be treated as the correct quantum description of the models under consideration? In this connection we mention that a serious problem arises with Ostrogradsky/Dirac/Horowitz as well as with BL formalisms when considering Gauss-Bonnet-Dilaton induced action. To be specific, in Subsection \ref{sec2.3} if $\gamma(\phi)$ is treated as a constant, then the contribution of Gauss-Bonnet term disappears from the Hamiltonian \eqref{M3}, and it reduces to \eqref{M1}. Indeed, it should since as mentioned, Gauss-Bonnet term is topologically invariant in $4$-dimensions, and so without functional coupling, it does not contribute to the field equations and the Hamiltonian as well. On the contrary, a constant $\gamma$ does not affect the form of the Hamiltonian \eqref{H1BL}, and it does not reduce to \eqref{hlb}. This means, if we had started with a constant $\gamma$ from the very beginning, all the terms appearing with $\gamma$ in \eqref{H1BL} would have been absent, and the end result would be \eqref{hlb}. While, after constructing the Hamiltonian with arbitrary $\gamma = \gamma(\phi)$, if we set it equal to a constant, then its contribution remains present, and we obtain a different Hamiltonian, altogether. Clearly this is wrong. Hence, we realize that boundary terms indeed play a crucial role while constructing the phase-space structure of non-linear theories. In fact, if boundary terms are taken into account from the very beginning, treating $h_{ij}$ as the basic variable, then Horowitz's formalism reduces to the MHF, as already demonstrated. It was also noticed that if Dirac algorithm is applied after integrating the action by parts, then it also yields Hamiltonian identical to MHF \cite{As13}. It is therefore suggestive to test the same for BL formalism too. In this section we shall first integrate actions by parts to get rid of the total derivative terms and follow the BL formalism thereafter, to explore the outcome.

\subsection{Scalar-tensor theory: minimal coupling;}

Upon integrating the action \eqref{A121} by parts, we obtain

\be\label{AI1} A_1 = \int\left[-{3\alpha\dot z^2\over 2N\sqrt z} +6\alpha k N\sqrt{z} +{9\beta\over N\sqrt{z}}\left\{\left({\ddot z\over N} - {\dot N\dot z\over N^2}\right)^2 + {2k\dot z^2\over z} + 4k^2 N^2\right\} + z^{3\over 2}\left({\dot\phi^2\over 2N} - NV\right)\right] dt.\ee
Replacing $\dot z$ by ${2N\over 3}q$ in view of \eqref{QBL}, the above action may be cast as,

\be\label{AI11} A_{1q} = \int\left[-{2\over 3} \alpha N {q^2\over \sqrt z} +6\alpha k N \sqrt{z} + {9\beta\over N\sqrt z}\left({4\over 9}\dot q^2 + {8 k N^2 q^2\over 9z} + 4 k^2 N^2\right) + z^{3\over 2}\left({\dot\phi^2\over 2N} - NV\right)\right] dt.\ee
Note that the action \eqref{AI11} cannot be expressed only in terms of velocities, due to the explicit presence of $q$ unlike \eqref{AQa}. However, similar situation arrived at, in the case of Gauss-Bonnet-Dilaton case, and so it doesn't matter. The canonical momenta are the following:

\be \label{mom1}p_q = {8\beta\over N\sqrt z}\dot q;~~p_\phi = {z^{3\over 2}\over N}\dot \phi;~~p_z = 0 = p_N.\ee
Dirac constraint analysis appears to be inevitable, since the action is singular. However as mentioned, the lapse function $N$ being the Lagrange multiplier, the constraint strongly vanishes, so that one can ignore it without loss of generality. Still, another primary constraint $p_z = 0$ is apparent. Nonetheless, as already noticed, in BL formalism, Dirac analysis may be bypassed despite the presence of the constraint $p_z = 0$ in the following manner. The Lagrangian density is:

\be L_{1q} = -{2\over 3} \alpha N {q^2\over \sqrt z} +6\alpha k N \sqrt{z}+ {9\beta\over N\sqrt z}\left({4\over 9}\dot q^2 + {8 k N^2 q^2\over 9z} + 4 k^2 N^2\right) + z^{3\over 2}\left({\dot\phi^2\over 2N} - NV\right),\ee
and hence the Hamiltonian reads as,

\be \label{H1}\begin{split}N\mathrm{H}^m_{MBL}& = p_q \dot q + p_z \dot z + p_\phi\dot\phi - L{1q}\\&
={N\sqrt z\over 16\beta}p_q^2 + {2\over 3}Nq p_z
+ {N\over 2 z^{3\over 2}}p_\phi^2 + {2\alpha Nq^2\over 3\sqrt z} - 6\alpha k N\sqrt z -  {8\beta kN q^2\over z^{3\over 2}} - {36\beta k^2 N\over \sqrt z} + NV z^{3\over 2},\end{split}\ee
where we have used \eqref{mom1} and replaced $\dot z$ by ${2N\over 3}q$, in view of \eqref{QBL}, and the suffix \{MBL\} now stands for `Modified Buchbinder-Lyakhovich' formalism. Finally as before, for the sake of comparison, if we perform the canonical transformation $q \rightarrow {3\over 2}x,~~\mathrm{and}~~p_q\rightarrow {2\over 3}p_x$, then the above Hamiltonian \eqref{H1} may be expressed in the following form,

\be\label{H2} \mathrm{H}^m_{MBL} = xp_z + {\sqrt z\over 36\beta}p_x^2 + {p_\phi^2\over 2 z^{3\over 2}}+{3\alpha\over 2\sqrt z}(x^2-4kz) - {18\beta k\over z^{3\over 2}}(x^2+2kz) + V z^{3\over 2},\ee
which is identical to $\mathcal{H}^m_{MHF}$ presented in \eqref{M1}.

\subsection{Scalar-tensor theory: non-minimal coupling;}

Here again, upon integrating the action \eqref{A121} by parts we obtain,

\be\label{AI2} A_2 = \int\left[-{3f\dot z^2\over 2N\sqrt z} -{3f'\dot\phi\dot{z}\sqrt z\over N} +6 f k N\sqrt{z} +{9\beta\over N\sqrt{z}}\Big\{\Big({\ddot z\over N} - {\dot N\dot z\over N^2}\Big)^2 + {2k\dot z^2\over z} + 4k^2 N^2\Big\} + z^{3\over 2}\Big({\dot\phi^2\over 2N} - NV\Big)\right] dt.\ee
Now, replacing $\dot z$ by ${2N\over 3}q$  in view of \eqref{QBL}, the above action \eqref{AI2} may be cast as,

\be\label{AI21}A_2 = \int\left[-{2\over 3} f N {q^2\over \sqrt z} -2 f' \sqrt {z} q\dot \phi + 6f k N \sqrt{z} + {9\beta\over N\sqrt z}\left({4\over 9}\dot q^2 + {8 k N^2 q^2\over 9z} + 4 k^2 N^2\right) + z^{3\over 2}\Big({\dot\phi^2\over 2N} - NV\Big)\right] dt.\ee
Canonical momenta may therefore be found as,

\be\label{mom2} p_q = {8\beta\over N \sqrt z}\dot q,~~p_\phi = -2f'q\sqrt z + {z^{3\over 2}\over N}\dot\phi,~~p_N = 0 = p_z.\ee
As before, leaving out the constraint associate with the lapse function, and replacing $\dot z = {2N\over 3} q$ in view of \eqref{QBL}, the Hamiltonian may be cast as,

\be\label{H3}\begin{split}N{\mathrm{H}^{nm}}_{MBL}& = p_q \dot q + p_z \dot z + p_\phi\dot\phi - L\\&
=N\left[{\sqrt z\over 16\beta}p_q^2 + {2\over 3}q p_z
+ {p_\phi^2\over 2 z^{3\over 2}} + {2f'\over z}q p_\phi + {2 f q^2\over 3\sqrt z} + {2f'^2 q^2\over \sqrt z} - 6kf\sqrt z -  {8\beta k q^2\over z^{3\over 2}} - {36\beta k^2 \over \sqrt z} + V z^{3\over 2}\right].\end{split}\ee
Finally, applying the canonical transformation relations $q \rightarrow {3\over 2}x,~\mathrm{and}~ p_q \rightarrow {2\over 3}p_x$, we obtain

\be\label{H4}\begin{split}{\mathrm{H}^{nm}}_{MBL}& = xp_z + {\sqrt z\over 36\beta}p_x^2 + {p_\phi^2\over 2 z^{3\over 2}} + {3x\over z}f'p_\phi + {3f\over 2\sqrt z}(x^2-4kz) - {18\beta k\over z^{3\over 2}}(x^2+2kz)+{9x^2 f'^2\over 2\sqrt z} + V z^{3\over 2}.\end{split}\ee
Clearly, ${\mathrm{H}^{nm}}_{MBL} \cong {\mathcal{H}^{nm}}_{MHF}$ presented in \eqref{M2}.

\subsection{Einstein-Gauss-Bonnet-Dilatonic action}

Eventually, in order to construct the correct Hamiltonian in connection with the Einstein-Gauss-Bonnet-Dilatonic action \eqref{A12}, let us integrate it by parts to obtain,

\be\begin{split}\label{AI3}A_3 =& \int\Bigg[\alpha\Big(-{3\dot z^2\over 2N\sqrt z}+6kN\sqrt z\Big) + {9\beta\over N\sqrt{z}}\Big\{\Big({\ddot z\over N} - {\dot N\dot z\over N^2}\Big)^2 + {2k\dot z^2\over z} + 4k^2 N^2\Big\} \\& -{\gamma'(\phi)\dot z\dot\phi\over N\sqrt z}\Big({\dot z^2\over N^2 z} + 12k\Big) + z^{3\over 2}\Big({\dot\phi^2\over 2N} - NV\Big)\Bigg] dt.\end{split}\ee

As before, replacing $\dot z$ by ${2N\over 3}q$ in view of \eqref{QBL}, the above action may be cast as,

\be\begin{split}\label{AI31}A_{3q} = &\int\Bigg[-{2\over 3} \alpha N {q^2\over \sqrt z}  + 6\alpha k N \sqrt{z} + {9\beta\over N\sqrt z}\left({4\over 9}\dot q^2 + {8 k N^2 q^2\over 9z} + 4 k^2 N^2\right)\\& -{2q\gamma'(\phi)\dot\phi\over 3\sqrt z}\Big({4q^2\over9 z}+ 12k\Big)+ z^{3\over 2}\left({\dot\phi^2\over 2N} - NV\right)\Bigg] dt.\end{split}\ee
Canonical momenta are now found as,

\be\label{mom3} p_q = {8\beta\over N \sqrt z}\dot q,~~p_\phi = - {2q\gamma'(\phi)\over 3\sqrt z}\Big({4q^2\over9 z}+ 12k\Big) + {z^{3\over 2}\over N}\dot\phi,~~p_N = 0 = p_z.\ee
As always, leaving out the constraint associated with the lapse function, and replacing $\dot z = {2N\over 3} q$ in view of \eqref{QBL}, the Hamiltonian may be cast as,

\be\label{H5}\begin{split}N{\mathrm{H}^{GB}}_{MBL} = &p_q \dot q + p_z \dot z + p_\phi\dot\phi - L\\&
=N\Bigg[{\sqrt z\over 16\beta}p_q^2 + {2\over 3}q p_z
+ {p_\phi^2\over 2 z^{3\over 2}}  + {2 \alpha q^2\over 3\sqrt z}  - 6k\alpha\sqrt z -  {8\beta k q^2\over z^{3\over 2}} - {36\beta k^2 \over \sqrt z}\\&+\frac{2q\gamma'(\phi)p_\phi}{3z^2}\left(\frac{4q^2}{9z}+12k\right)+\frac{2q^2\gamma'^2(\phi)}{9z^{5\over 2}}\left(\frac{4q^2}{9z}+12k\right)^2 + V z^{3\over 2}\Bigg],\end{split}\ee
Finally, the set of canonical transformations $q \rightarrow {3\over 2}x,~\mathrm{and}~ p_q \rightarrow {2\over 3}p_x$, allows one to express the Hamiltonian \eqref{H5} as,

\be\label{H6}\begin{split}{\mathrm{H}^{GB}}_{MBL}= &x p_z + {\sqrt z\over 36\beta}p_x^2 + {p_\phi^2\over 2 z^{3\over 2}}  + {3\alpha\over 2\sqrt z}(x^2-4kz) - {18\beta k\over z^{3\over2}}(x^2+2kz)\\&+\frac{x\gamma'(\phi)}{z^2}\left(\frac{x^2}{z}+12k\right)p_\phi+\frac{\gamma'^2(\phi)x^2}{2z^{5\over 2}}\left(\frac{x^2}{z}+12k\right)^2 + V z^{3\over 2}.\end{split}\ee
As a result one finds, ${\mathrm{H}^{GB}}_{MBL} \cong {\mathcal{H}^{GB}}_{MHF}$ presented in \eqref{M3}. It is important to note that in the process of constructing the Hamiltonian starting from a divergent free action, the pathology discussed in regard of canonical formulation of Einstein-Gauss-Bonnet-Dilatonic action in the presence of higher-order term is also removed.

\section{Application}

It is mentioned in the introduction that canonical formulation is a precursor to canonical quantization. In the absence of a viable quantum theory of gravity, it is suggestive to canonically quantize the cosmological equation and study quantum cosmology to extract some ethos of pre-Planck era. For example, one can explore the Euclidean wormhole solution. Nonetheless, `cosmological inflationary scenario' has been developed since 1980, to solve horizon, flatness (fine tuning), structure formation and monopole problems, singlehandedly. Short-lived $(10^{-36} - 10^{-26})$s. inflation, occurred just after Planck's era and falls within the periphery of `quantum field theory in curved space-time'. To be more specific, `inflation is a quantum theory of perturbations on the top of the classical background', so that the energy scale of the background remains much below Planck's scale. Nonetheless in this context, Hartle \cite{Hartle} prescribed that, most of the important physics may still be extracted from the classical action provided, the semiclassical wave-function is strongly peaked. The reason being, in that case correlation between the geometrical and matter degrees of freedom is established, and hence the emergence of classical trajectories (i.e. the universe) is expected. Hence, quantization and an appropriate semiclassical approximation must be treated as a forerunner to study inflation. \\

Canonical quantization and the semiclassical wave-function in connection with the Hamiltonian \eqref{H4} for non-minimally coupled higher order theory had been presented in \cite{As12}, which reduces to the minimally coupled case when the coupling parameter becomes constant \cite{As3}. The Hamiltonian operator was found to be hermitian, standard probabilistic interpretation holds, and the semiclassical wave-functions was found to be oscillatory about the classical inflationary solution. Inflation has been studied and the parameters are found with excellent agreement with the observational constraints \cite{Planck1, Planck2}. Gravitational perturbation has also been studied. \\

In \cite{As16} again, the quantum counterpart of the Hamiltonian \eqref{H6} in connection with Einstein-Gauss-Bonnet-Dilatonic coupled action has been presented. Hermiticity of the Hamiltonian operator has been established, probabilistic interpretation is explored, and the semiclassical wave-function is found to be oscillatory about a classical inflationary solution. Finally, we have studied inflation and found that the inflationary parameters more-or-less satisfy observational constraints \cite{Planck1, Planck2}. In a nut-shell, the results obtained in \cite{As16} are the following.

\be \label{qh2} \begin{split} {i\hbar}\frac{\partial\Psi}{\partial \sigma}&=\left[-\frac{\hbar^2 \phi}{54\beta_0 x}\left(\frac{\partial^2}{\partial x^2} +\frac{n}{x}\frac{\partial}{\partial x}\right) -\frac{\hbar^2}{3x\sigma^{\frac{4}{3}}}\frac{\partial^2}{\partial \phi^2}+\frac{2i\hbar \alpha_0} {\sigma}\left(\frac{1}{\phi^2}\frac{\partial}{\partial \phi}-\frac{1}{\phi^3} \right) - \frac{2i\hbar \gamma_0 x^2}{3\sigma^{7\over 3}}\left(2\phi{\partial\over \partial \phi} + 1\right) +V_e\right]\Psi\\&=\widehat H_e \Psi, \end{split}\ee
where, the proper volume, $\sigma=z^{\frac{3}{2}} = a^3$ plays the role of internal time parameter, and $n$ is the operator ordering index. In the above equation,$\widehat H_e$ is the effective hermitian Hamiltonian operator, while the  the effective potential $V_e$ is given by,

\begin{center}
    \be V_e = \frac{3\alpha_0^2 x}{\sigma^{2\over 3}\phi^4} - \frac{4\alpha_0\gamma_0 x^3}{\sigma^2 \phi} + \frac{4\gamma_0^2 x^5\phi^2}{3\sigma^{10\over 3}} + \frac{\alpha_0 x}{\sigma^{2\over 3}\phi} + \frac{\lambda^2\sigma^{2\over 3}\phi^2}{3x}+{2\sigma^{2\over 3}\Lambda M^2_P\over x}. \ee
\end{center}
The effective Hamiltonian operator is found to be hermitian for $n = -1$, which selects the operator ordering parameter from physical consideration. Standard quantum mechanical probability interpretation also holds. Under a suitable (WKB) semiclassical approximation, the wave-function has been found to be,

\be\label{psif} \Psi = \Psi_0 e^{\frac{i}{\hbar}\left[ -\frac{6\alpha_0\lambda z^2}{a_0\phi_0}+16\gamma_{0}a_{0}^2\phi_{0}^2\lambda^3\sqrt{z}\right]},\ee\\
which exhibits oscillatory behaviour about the classical inflationary solution $a = a_0 e^{\lambda t}$, where, $\alpha_0, \phi_0, \gamma_0$ are constants. We have also presented several sets of inflationary parameters in \cite{As16}, which depict that the spectral index of scalar perturbation and the scalar to tensor ratio lie within the range $0.967 \le n_s \le 0.979$ and $0.056\le r \le 0.089$ respectively, showing reasonably good agreement with the recently released data \cite{Planck1, Planck2}. The number of e-folding also remains within the acceptable range $46 < \mathrm{N} < 73$, which is sufficient to solve the horizon and flatness problems.

\section{Concluding remarks}

Although initiated two centuries back, canonical formulation of higher-order theory of gravity is particularly non-trivial. In fact, only after probing Dilatonic coupled Gauss-Bonnet action, it is learnt that divergent terms play a vital role to formulate correct quantum dynamics of non-linear gravity theory. The scheme is therefore first, to express the action in terms of the basic variable $h_{ij}$, otherwise if expressed in terms of the scale factor, as commonly done, some unwanted divergent terms are removed in the process of integration by parts, which are unaccredited by the variational principle. Next, unless divergent terms are taken care of, the Hamiltonian is found to be different, which is related through canonical transformation though, such transformation cannot be carried over in the quantum domain due to non-linearity. It is shown that in the case of Einstein-Gauss-Bonnet-Dilatonic coupled action in $4$-dimension that, unless the action is divergent free, an erroneous Hamiltonian is constructed, since it does not reflect the topological invariance of the theory. This proves the importance of divergent terms in higher order theories. In this respect the difference of BL formalism with MHF is apparent. In fact BL formalism produces identical Hamiltonian as obtained earlier following Ostrogradsky's, Dirac's or Horowitz's formalisms. However, MHF is essentially the Horowitz formalism, after expressing the action in terms of the three space curvature and taking care of the total derivative terms under integration by parts. It was shown that following the same route if Dirac's algorithm is applied, the Hamiltonian becomes identical to the one found following MHF, and one obtains unique quantum description. Here, we reveal that the same is true with BL formalism. In fact, BL formalism not only bypasses constraint analysis, as in the case of Horowitz's formalism, it also does not require auxiliary variable to cast the action in canonical form, which is a bit intricate. In a straightforward manner, it establishes diffeomorphic invariance, and therefore is the easiest technique to handle higher-order theories.\\

\end{document}